\documentclass[11pt]{article}
\usepackage{graphicx}

\begin{document}

\pagestyle{plain} \setcounter{page}{190}

{\large\bf Balmer jump variations of HD 215441 (Babcock's star)\\[5.0mm]}

{\bf N. A. Sokolov}\\

{\scriptsize Central Astronomical Observatory at Pulkovo,
St. Petersburg 196140, Russia E-mail: sna@gaoran.spb.su}\\[5.0mm]

{\bf Abstract.} Balmer jump discontinuity of the peculiar star
HD 215441 was measured on 14 continuous energy distributions
obtained by S. J. Adelman at Kitt Peak National and Palomar Observatories.
The results show that the Balmer jump vary by about 0.05 dex
over the cycle of the star. A comparison with new Kurucz model atmospheres
shown that the Balmer jump for the phase 0.5 corresponds to a 14750 K,
log(g)=3.0 and for the phase 1.0 corresponds to a 15750 K, log(g)=3.0.

\section{Introduction}

Photometric variations in peculiar A (Ap) stars due to variable
flux redistribution from the ultraviolet to the visual region of the
spectrum. In this model, the variation of line or continuum opacity
over the surface of the star produces, as the star rotates, a variation
in line blanketing in the ultraviolet and hence in the amount of flux
redistributed. A natural consequence of this model is that a star will
exhibit photometric and spectrum variations in phase and with the same
period.

Leckrone (1974) has presented evidence from ultraviolet
photometry and Adelman (1983) has presented evidence from optical
region spectrophotometry for variable flux redistribution in the
magnetic Ap star HD 215441. This star is one of the outstanding Ap star
photometric variables with amplitudes of 0.$^m$11, 0.$^m$14
and 0.$^m$19 in V, B and U
respectively (Stepien, 1968). The surface magnetic field oscillates
between 32 and 35 kG in phase with the UBV variations (Preston, 1969).
Leckrone obtained ultraviolet photometry between 1550~\AA~and 4250~\AA.
He found that at the wavelength 2980~\AA~the variability is in phase
with the optical region. On the other hand, Adelman found that around
the cycle of variability the Balmer jump discontinuity (BD) exhibits
changes and correlate with the feature at $\lambda$~5200~\AA, but he
could not measure the size of the Balmer jumps for this star.

In this work we will derive the Balmer jump discontinuity for
the magnetic Ap star HD 215441. We will also reconcile the variations of BD
with the conclusions of other authors about photometric variations for
this star.

\section{Source of the data and reduction}

A series of optical energy distributions of the magnetic Ap
star HD 215441 obtained by S. J. Adelman at Kitt Peak National and
Palomar Observatories were made available to the authors through
Strasburg Data Center by using Internet. These data also have been
published by Adelman (1983). The total number of scans was
thirty-eight, but such as the period is rather long and the star
relatively faint, these scans obtained on the same night were averaged
together. The final number of the optical energy distributions was
fourteen.

\begin{figure*}
\centerline{\includegraphics[width=80mm, angle=0]{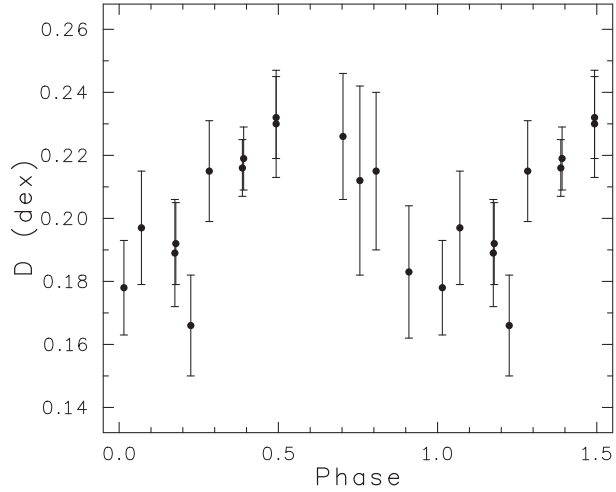}}
\caption{The HD 215441 size of the Bailer jump as a function of phase.}
\end{figure*}

The use of UBV and/or {\it uvby} photometry to deduce the
amount of interstellar reddening has been questioned. If one has
sufficient ultraviolet photometry especially in the region near
$\lambda$~2200~\AA, then one might have a way to solve this problem,
but HD 215441 was not observed by ANS (van Dijk et al., 1978) and
Leckrone's (1974) values also do not cover this region. On the other
hand, changing the reddening does not markedly change the size of
the Balmer jump. This fact allows us to correct all data for the interstellar
reddening using the values E(B-V)=0.$^m$21 obtained by Peterson (1969).

We have computed the size of the Balmer jump by extrapolating
the two fitted curves to $\lambda$=3700~\AA~separately in the Balmer
continuum and Paschen continuum as described by Sokolov (1995). As far
as, the errors of BD are concerned, we computed them by taking into
account the errors of the two fitted curves.

The rotational phase of the optical energy distributions was
computed according to the ephemeries given by Leckrone (1974).

\bigskip
JD(max. light $\lambda$ $>$ 3300 \AA) = 2436864.88$^d$ +
 9.4871$^d$~{\it E}
\bigskip

The phase of maximum light ($\Phi$ = 0.0) coincides with the phase
of maximum magnetic effective field (Borra and Landstreet, 1978).

\section{Results and discussion}

We applied the method of determination of the size of the
Balmer jump (Sokolov, 1995) for 14 continuous energy
distributions for the magnetic Ap star HD 215441. In the Figure 1
we can see that BD strongly depend on the rotational phase and show
a sinusoidal variation. The variation of BD is in antiphase to the
surface magnetic field (see Preston, 1969) and to the effective
magnetic field (see Borra and Landstreet, 1978). A comparison with
UBV photometry show that the BD is smallest when the Paschen continuum
the bluest according to B-V (Stepien, 1968). On the other hand, the
hydrogen line equivalent widths are bigger when the BD is smallest
(Malanushenko et al., 1992).

The Balmer jump discontinuity is one of the key parameters
for stellar spectra. Moreover, the Balmer jump is used for stellar
classification of the main sequence stars, because there is very good
correlation with effective temperature. However, Krautter (1977)
claims that the effective temperature is constant over the rotation
cycle of Bacock's star HD 215441, because the amplitudes of the
equivalent width variations did not depend on the excitation potential.
His argument supports the earlier analysis of the flux variation
over a wide spectral range by Leckrone (1974). But, recently,
Malanushenko et al. (1992) were found a decreasing amplitude for an
increasing order of the Balmer lines. They were also computed
the effective temperature from the comparison equivalent line widths
with Kurucz model atmospheres. Their equivalent widths would require
a low log(g) = 3.0 and a high temperature of about 18 000 K.
Moreover, Adelman (1983) has found that the immediate Balmer jump region
is fit by the predictions of a 18250 K, log(g) = 4.0 model for the
phase 0.93 and by those of a 16250 K, log(g) = 4.0 model for the phase 0.51.
Compared to the new Kurucz model atmospheres (Kurucz, 1993),
our measured the size of the Balmer jumps shown the predictions of
a 14750 K, log(g) = 3.0 for the phase 0.5 and those of a 15750 K,
log(g) = 3.0 for the phase 1.0 with metallicity [M/H]=1.0 which is
in the agreement with the temperature of Babcock's star (Adelman, 1983).
However, using Kurucz models which are not designed to simulate
the complex nature of magnetic star atmospheres is questionable. The
strong magnetic surface field is a prime candidate for causing a severe
deviation of the data relative to standard model atmospheres and is
not taken into account by Kurucz.

\bigskip
{\noindent\large References}

\bigskip
\begin{description}

\item[] {Adelman, S. 1983, A\&AS 51, 551}
\item[] {Borra, E., Landstreet, J. 1978, ApJ 222, 226}
\item[] {Krautter, A. 1977, ApJ 216, 33}
\item[] {Kurucz, R. L. 1993, data on magnetic tape}
\item[] {Leckrone, D. 1974, ApJ 190, 319}
\item[] {Manalushenko, V., Polosukhina, N., Weiss, W. 1992, A\&A 259, 567}
\item[] {Preston, G. W. 1969, ApJ 156, 967}
\item[] {Sokolov, N. 1995, A\&AS 110, 553}
\item[] {Stepien, K 1968, ApJ 154, 945}
\item[] {van Dijk, W., Kerssies, A., Hammerschlag-Hensberge, G.,
 Wesselius, P. R. 1978, A\&A 66, 187}

\end{description}
\end{document}